\documentclass[12pt]{article}
\usepackage{aaspp4}
\begin{document}

\def\kms{$\rm {km}~\rm s^{-1}$}
\def\ni{\noindent}
\def\msun{{\rm M}_\odot}
\def\deg{\ifmmode^\circ\else$^\circ$\fi}
\def\etal{{\it et~al.}}
\def\ie{i.e.,}
\def\Mtil{\ensuremath{\tilde{M}}}
\def\Meff{$\tilde{M}_{eff}$}
\def\im{iM_{2}}
\title{Toward a Symmetrization of Gravity I. Classical Formulation: Results and
Predictions; Dark Matter With a Difference}
\author{Terry Matilsky} 
\affil{Department of Physics and Astronomy, Rutgers, The State
University of New Jersey\\ 136 Frelinghuysen Rd., Piscataway, NJ 08854-8019}
\affil{matilsky@physics.rutgers.edu}

\begin{abstract}
We propose an additional term in the classical gravitational force law, 
which is repelling in nature, and which may solve the dark matter problem.   As an
inverse cube field interaction, it operates over 4 real spatial dimensions
and its effect on our observable 3-D space may account both for flat
rotation curves and 
standard Newtonian dynamics at small radial distances. 
By utilizing cosmological clustering scales, 
we can derive the universal interaction strength, and show that this naturally 
leads to an altered Planck mass that becomes the neutron/proton rest mass. 
Moreover, the correct value for the electron rest mass can be predicted using 
only classical electrostatics coupled with the current theory. On cosmological scales, the interaction easily accounts for the acceleration of the Hubble flow.
\end{abstract}

\section{Introduction}

For well over 50 years astrophysicists have been struggling with the
apparent mass discrepancies that seem to exist in large scale structures
throughout the universe (Zwicky 1937). The discovery of asymptotically
flat galactic rotation curves at large distances from the core (Bosma 1978) started an
avalanche of ``dark" matter (DM) ideas in an effort to explain the growing body of data. Interestingly, it seems that many current
versions of DM run into serious difficulties when compared with observed
properties of galaxies (Sellwood 2000). Moreover, with the discovery of the
likelihood of a global acceleration to the Hubble flow (Perlmutter \etal\ 1999;
Riess \etal\ 1998), it would seem that the time is ripe for a fundamentally
new approach, one which might have the possibility of explaining many
disparate observational problems. To this end, we postulate an additional mode for gravitational interaction in which the 
field associated with a hypothetical particle is everywhere repelling.  For simplicity, 
we call the mass associated with this field $\tilde{M}$, to distinguish it from
``ordinary"  
matter, M. If the repelling field is inverse-square in nature, the combined gravitational and anti-gravitational force between
the M and $\Mtil$\ will be independent of separation, and thus 
merely implies a change in the magnitude of G.

However, a potential of the form $\Phi_{\Mtil} \propto  \frac{1}{r^{2}}$ has several 
interesting aspects that we will investigate here:

A)  Although repelling by nature, it can substantially enhance the gravitational acceleration in the
limit of large distances from the center of the potential, as well as yield  standard Newtonian dynamics as r $\rightarrow$ 0.

B)  It suggests a force acting over a 4 dimensional, real spatial manifold. 

An obvious characteristic of the superposition of an inverse-cube repelling
force with an inverse-square attracting force is that it establishes and
requires a length scale to be invoked. Only at one particular separation of an $M-\Mtil$\ pair will the magnitudes of the field strengths be equal. Thus, if this
separation of an isolated M-$\tilde{M}$ pair is given as $R_{0}$, any perturbation of
either particle tends toward equilibrium.  (By equilibrium, we mean
here that the absolute value of the two fields are equal).  If the separation
is increased, the inverse-square attractive term dominates (thereby decreasing the
separation ), while if the separation is decreased, the
inverse-cube repelling term dominates  (thereby sending the system back toward
larger values of the separation). 

At equilibrium, we have:
\begin{equation}
\tilde{g} \equiv \left|\frac{\tilde{G}\tilde{M}}{R^{3}_{0}} \right| = \left|\frac{GM}{R^{2}_{0}} \right|
\equiv g
\end{equation}
So, for the case of M=$\tilde{M}$, it follows that $\frac{\tilde{G}}
{G} = R_{0}$.
Since we invoke ``equilibrium'' (M=$\tilde{M}$) at the limit of large R, it seems
natural to choose $R_{o} \approx 100$ Mpc, which is the putative lower bound of
true global homogeneity (Tucker, Lin \& Shectman 1999).  With this value,

\begin{equation}
\frac{\tilde{G}}{G} \approx 3 \cdot 10^{26} cm \ \ \mbox{and thus} \ \ 
\tilde{G} \approx 2 \cdot 10^{19} cgs \ \  \mbox{units}
\end{equation}

Note that this distance scale is merely a currently observed quantity, and as such, does not \emph{in principle}  depend on the time history of the Universe.  If the Universe is flat or open, this scale length can exist at all epochs. Thus, we don't necessarily need to invoke time variability of fundamental constants.  If the Universe is closed, there will be an epoch where the scale factor implies global distances smaller than 100 Mpc, but this still does not rule out anything that follows here.
In any event, we bear in mind that this value is just a working hypothesis, and  has no real effect on our conclusions for dynamics, since the \emph{product} of $\tilde{G}$ and \Mtil\   is all that appears in every  calculation.  
	However, we shall see in section 3 that our ``guess" is apparently quite close to a value that redefines the Planck scales in a strikingly reasonable way, and also (in section 4) that this provides  a natural value of the density of \Mtil\  to explain the apparent acceleration of the Hubble flow.  In this regard, it is useful to point out, that as a repelling field, it will will be smooth over large distances, being characterized as a fluid with some density $\tau$.  We consider this idea below.

\section{Dynamics in the Fourth Dimension}
To postulate a force that varies as $\frac{1}{r^{3}}$ suggests that it
operates in four spatial dimensions. In a universe devoid of any  clustering 
of ordinary matter, such a repelling system would be modelled as a uniform, continuous fluid, obeying Birkhoff's theorem for four dimensions.  When the ordinary matter, M, begins to cluster, it attracts the \Mtil.   If the \Mtil\ is compressible,
we can easily envision an equilibrium configuration whereby gravity is
completely neutralized by  density gradients set up by the \Mtil. However, if the \Mtil\ fluid has a maximum allowable density (or is incompressible), it can now yield localized regions that can augment the gravitational field due to the M alaone.  For example, it is possible for  a localized region of self-repelling \Mtil\  to form a shell surrounding the real M, attracted toward the center of the potential well by the Newtonian field, and repelled by the bulk of the \Mtil\ in the shell.
	The problem with this picture is that it is difficult to see how, in a 4-dimensional manifold, one can have an inverse cube law for one interaction, and an inverse square law for another. We point out  the recent finding by Randall and Sundrum (2000), which claims that the standard experimental results for gravity (such as obedience to a 3 dimensional Poisson equation) is consistent with even an \emph{infinite} 4th spatial dimension, if the component of the metric in the new dimension depends on the coordinates in that dimension.
Without going into detail here, we merely accept this as a possibility and explore the phenomenological consequences of our postulated interaction.
  
	Returning to the simple configuration of a shell of \Mtil\ material as outlined above, we can evaluate the potential at any point P in the shelll's interior:  $\phi_{\Mtil} = \int \frac{\tilde{G}d\Mtil}{u^2}$, where $u$ is the distance from a point on the shell with mass element $d\Mtil$\ to to the point P.   The
integration is straightforward and we find:

\begin{equation}
\phi_{\Mtil} = 2\pi \tilde{G}\sigma \frac{R}{r} ln \frac{R+r}{R-r}
\end{equation}
where R is the radius of the \Mtil\ shell, $\sigma$ is its surface density on the 3-sphere, and r is the radial displacement of P from the center of the shell.  We immediately
find the field:
\begin{equation}
|\tilde{g}(r)| = \frac{ 2 \pi \tilde{G}\sigma R}{r} \left[\frac{1}{r} ln \frac{R+r}
{R-r} - \frac{2R}{R^{2} - r^{2}} \right]
\end{equation}
directed radially inward.

For r$\ll$R, expanding the logarithm yields the following result for the
acceleration:

\begin{equation}
\tilde{g}(r) = - \frac{8\pi}{3R^{2}} \tilde{G}\sigma \vec{r}\ + O(r^2)
\end{equation}

The \Mtil\ on the surface of the sphere  thus pushes inward differentially near M,  effectively augmenting  the gravitational potential due to the real mass alone.  Of course, the configuration surrounding a flattened system will be
different, but the above should be sufficient to demonstrate that this idea holds the possibility of solving the dark matter problem.  For a required velocity of $\sim
100\ km/sec$ at r=10 kpc, for example, we could have R=30 kpc with a total \Mtil\
 $\approx 10^{42}$\ gm in the shell.  It is clear that the \Mtil\ will not be confined to a shell, and ought to fill space between the galaxies, but the dominant portion of interaction will be from the regions closest to the galactic center, since the $\frac{1}{r^3}$ interaction will fall off rapidly as R gets  larger.
Moreover, as r $\rightarrow 0,\  \tilde{g}(r) \rightarrow 0$ , thus recovering a purely Newtonian field.

\section{The Interaction Strength and Implications For
Microphysics}

Since $\tilde{G}$ has a different dimensionality than G, we can now use it
instead of G, to form the standard units of m, r, and t by combining
$\tilde{G}$ with h and c.  So doing yields:

\begin{eqnarray}
\tilde{m} & = & (\frac{h^{2}}{\tilde{G}})^{1/3} = 1.4 \cdot 10^{-24} g
\nonumber \\ 
\tilde{r} & = & (\frac{\tilde{G}\cdot h}{C^{3}})^{1/3} = 1.8 \cdot 10^{-13} 
cm \nonumber \\
\tilde{t} & = & (\frac{\tilde{G}\cdot h}{c^{6}})^{1/3} = 6 \cdot 10^{-24} s
\end{eqnarray}
Note that although \emph{both} $\tilde{m}$ and $\tilde{r}$ agree quite
well with the classical 
definition of the nuclear mass and radius, the two are coupled together;
any value of G (or $\tilde{G}$) that yields the correct m will perforce
yield the corresponding r.  Nonetheless, the similarity of $\tilde{m}$ to
the observed nuclear mass is remarkable.

Another interesting result is obtained when the classical electrostatic force
is equated with the $\tilde{M}$ interaction at the distance $\tilde{r}$, using $\tilde{m}$ as one
of the masses,  
and solving for the other mass.  We get:

\begin{equation}
\frac{\tilde{G}\tilde{m}m_{x}}{\tilde{r}^{3}} = \frac{q_{1}q_{2}}{\tilde{r}^{2}}
\end{equation}
so
\begin{equation}
m_{x} = \frac{\tilde{r}q_{1}q_{2}}{\tilde{G}\tilde{m}} = \frac{1.8 \cdot 10^{-13} \cdot
(4.8 \cdot 10^{-10})^{2}}{2 \cdot 10^{19} \cdot 1.4 \cdot 10^{-24}} \approx
1 \cdot 10^{-27} g
\end{equation}

This, of course, is very close to the value for the electron's rest mass.  
Thus, we apparently have a way to uniquely determine both the proton and
electron rest masses.  While these results could be considered as merely chance
numerical coincidences, it is possible that this result could point toward some sort of exchange interaction, whereby quantum mechanical operators allow interchanges between fundamental M and \Mtil\ ``particles".

\section{The Hubble Flow and Cosmological Considerations}

	A repelling force has obvious applicability toward observational cosmology.  If we consider a dust filled universe, with no density gradients
in \(M\) or \Mtil, we ought to be able to approximate the observed Hubble flow in a simple fashion.  The repelling field on a shell of material, due to
the presence of \Mtil\ interior to the shell, will be:

\begin{equation}
\tilde{g}(r) =  \frac{\pi^{2}}{2} \tilde{G}\tau \vec{r}\ ,
\end{equation}
where $\tau$\ is the 4-dimensional volume density of the \Mtil. (This result follows from the fact that the hyper-volume of the sphere of \Mtil\ is: $\frac{\pi^{2}}{2}r^{4}$).  Thus, our interaction superficially has  the same effect as
the cosmological constant, whereby:
\begin{equation}
\frac{\pi^{2}}{2} \tilde{G} \tau = \frac{1}{3}\Lambda .
\end{equation}
	Using the results of Perlmutter \etal\ (1999) and
Riess \etal\  (1998), we set $\Omega_{\Lambda} \equiv \Lambda/3H_0^2=0.7$, and therefore find (with $H_0=70km/s/Mpc$) : $\tau \approx 10^{-57} g/cm^{4}$.
We are now afforded another consistency check with regard to our original
assumption of equal mass distributions of M and \Mtil\ over 100 Mpc.  Using the
4 dimensional volume for \Mtil, we find that \Mtil\ $\approx 10^{50}$\ g.  If we assume that the average 3-D density of real matter is given by $\rho \approx 3 \cdot 10^{-31}\ g/cm^{3}$ for the M (i.e. there is no
``dark" matter), we get $M \approx 3 \cdot 10^{49}$\ g. 
	However, since the cosmological principle implies that expansion proceeds in the same way for all shells, we see that $\tau / \rho \propto 1/a$, where $a$ is the scale factor for the expansion.  Thus, the repulsive stress is diluted over time (unlike the stress due to a $\Lambda$\ term), and therefore the acceleration may end at some time in the future, thereby avoiding the problems
that a vacuum dominated, eternally accelerating universe faces (see Barrow, Bean and Magueijo, 2000).

	We would like to thank Tad Pryor, Jerry Sellwood, Jim Peebles, John Conway, Arthur Kosowsky, Tina Kahniashvili 
and Stacy McGaugh for stimulating discussions, and Patty Gulyas for help in the manuscript preparation.

\end{document}